\documentclass[a4paper,fleqn]{cas-dc}
\usepackage{lineno}
\usepackage{graphicx}
\usepackage{float}
\usepackage{subcaption}
\usepackage[labelfont=rm]{caption}
\usepackage{capt-of}
\usepackage{placeins}
\usepackage{enumitem}

\usepackage[numbers,sort&compress]{natbib}
\setcitestyle{square,comma,numbers}

\usepackage{float}
\def\tsc#1{\csdef{#1}{\textsc{\lowercase{#1}}\xspace}}
\tsc{WGM}
\tsc{QE}
\tsc{EP}
\tsc{PMS}
\tsc{BEC}
\tsc{DE}

\begin{document}
\let\WriteBookmarks\relax
\def\floatpagepagefraction{1}
\def\textpagefraction{.001}

\shortauthors{P. Federičová et~al.}
\shorttitle{Summary of quality control (QC) of ATLAS18 production ITk strip
sensors}

\title [mode = title]{Summary of quality control (QC) of ATLAS18 production ITk strip
sensors}

%
\author[1]{P. Federičová}[orcid=0000-0003-4176-2768]

\ead{pavla.federicova@cern.ch}

\cormark[1]

\author[2]{A. Affolder}
\author[2]{K. Affolder}
\author[3]{A. Awais}
\author[3]{G.A. Beck}
\author[3]{A. J. Bevan}
\author[3]{Z. Chen}
\author[4]{J. Dandoy}
\author[3]{I. Dawson}
\author[2]{V. Fadeyev}
\author[5]{J. Fernandez-Tejero}
\author[6]{E.~C. Hill}
\author[7]{S. Hirose}
\author[8]{L. Hommels}
\author[8]{T. Ivison}
\author[4]{C. Jessiman}
\author[8]{K. Kariyapperuma}
\author[2]{S. Katznelson}
\author[4]{J. Keller}
\author[4]{C. T. Klein}
\author[4]{T. Koffas}
\author[9]{I. Kopsalis}
\author[1]{J. Kozáková}
\author[1]{J. Kroll}
\author[1]{M. Kůtová}
\author[1]{J. Kvasnička}
\author[7]{K. Maeyama}
\author[3]{R.R. Marcelo Gregorio}
\author[2]{F. Martinez-Mckinney}
\author[1]{M. Mike\v{s}t\'{i}kov\'{a}}
\author[3]{P. S. Miyagawa}
\author[2]{L. Morelos-Zaragoza}
\author[8]{K. Nakamura}
\author[2]{Q. Paddock}
\author[7]{K. Sato}
\author[4]{E. Staats}
\author[1]{P. Tůma}
\author[5]{M. Ullan}
\author[10]{Y. Unno}
\author[4]{Y. Zhao}
\author[3]{S. C. Zenz}

\affiliation[1]{organization={Institute of Physics, Czech Academy of Sciences},
    addressline={Na Slovance 2}, 
    city={Prague},  
    postcode={18200},    
    country={Czech Republic}}
\affiliation[2]{organization={Santa Cruz Institute for Particle Physics (SCIPP), University of California},
    city={Santa Cruz},
    postcode={CA 95064}, 
    country={USA}}
\affiliation[3]{organization={Particle Physics Research Centre, Queen Mary University of London},
    addressline={G.O. Jones Building, Mile End Road}, 
    state={London E1 4NS},
    country={United Kingdom}}
\affiliation[4]{organization={Physics Department, Carleton University},
    addressline={Physics Department, Carleton University}, 
    city={Ottawa},
    state={Ontario},
    postcode={K1S 5B6},
    country={Canada}}
\affiliation[5]{organization={Instituto de Microelectr{\'o}nica de Barcelona (IMB-CNM)},
    addressline={CSIC, Campus UAB-Bellaterra}, 
    city={Barcelona},
    postcode={08193},
    country={Spain}}
\affiliation[6]{organization={Department of Physics, University of Toronto},
    addressline={60 Saint George St.}, 
    city={Toronto},
    postcode={Ontario M5S1A7},
    country={Canada}} 
\affiliation[7]{organization={Institute of Pure and Applied Sciences, University of Tsukuba},
    addressline={1-1-1 Tennodai}, 
    city={Tsukuba},
    postcode={Ibaraki 305-8571},
    country={Japan}}
\affiliation[8]{organization={Cavendish Laboratory, University of Cambridge},
    addressline={JJ Thomson Avenue}, 
    city={ Cambridge},
    postcode={CB3 0HE},
    country={United Kingdom}}   
\affiliation[9]{organization={Department of Physics, National Technical University of Athens},
    addressline={9 Iroon Polytechniou St., Zografou Campus}, 
    city={ Athens},
    postcode={15780},
    country={Greece}}
\affiliation[10]{organization={Institute of Particle and Nuclear Study, High Energy Accelerator Research Organization (KEK)},
    addressline={1-1 Oho}, 
    city={Tsukuba},
    postcode={Ibaraki 305-0801},
    country={Japan}} 
    
\cortext[cor1]{Corresponding author.}

\begin{abstract}
To address the demanding operational requirements of the High-Luminosity upgrade of the
Large Hadron Collider (HL-LHC), the ATLAS experiment is replacing its current Inner
Detector with a~new all-silicon Inner Tracker (ITk). The ITk will feature an active sensor area
of 165 m², with its outer tracking layers populated by approximately 18,000 ATLAS18 n$^{+}$-in-p
silicon strip sensors. The silicon sensors, available in eight geometries tailored to two barrel and six endcap module types, respectively, are designed to tolerate fluences of up to 1.6 × 10$^{15}$ neq/cm² and ionizing doses of 66 Mrad. A~comprehensive, multi-year Quality Control (QC) program is underway across multiple international institutes to evaluate these ITk strip sensors for mechanical and electrical conformity. The QC process includes IV/CV characterization, full strip tests, long-term
current stability monitoring, visual inspection, and metrology tests. To manage the high
throughput of about 500 sensors per month, the collaboration has implemented standardized
test procedures, software packages for data monitoring and integrity checks, unified data formats, and automated analysis tools. The standardization ensures consistent pass/fail evaluation and centralized data handling that enables effective identification of trends and anomalies at all sites during multi-year production.

This contribution presents an overview of the ITk strip sensor production and QC
framework, along with key findings throughout the whole production, such as charge-up of
sensors, stability of the leakage currents, nonrecoverable IV breakdown, and low inter-strip
isolation within wafers. It provides insights into sensor yield, quality trends, and reviews
specific case studies, such as p-stop doping non-uniformity. Over 91\% of the production,
totaling over 590 batches, were tested and accepted as-is. Six batches were rejected: two due
to instability and non-recoverable IV breakdown found in QC, and four due to non-uniform p-stop
doping found during Quality Assurance (QA) testing. These account for 2.8\% of the total tested sensors. Additionally, 1.8\% of individual
sensors are rejected after visual inspection, non-recoverable IV breakdown, and other
issues.

\end{abstract}



\maketitle

\section{Introduction}
\begin{figure*}
	\centering
		\includegraphics[width=0.9\textwidth]{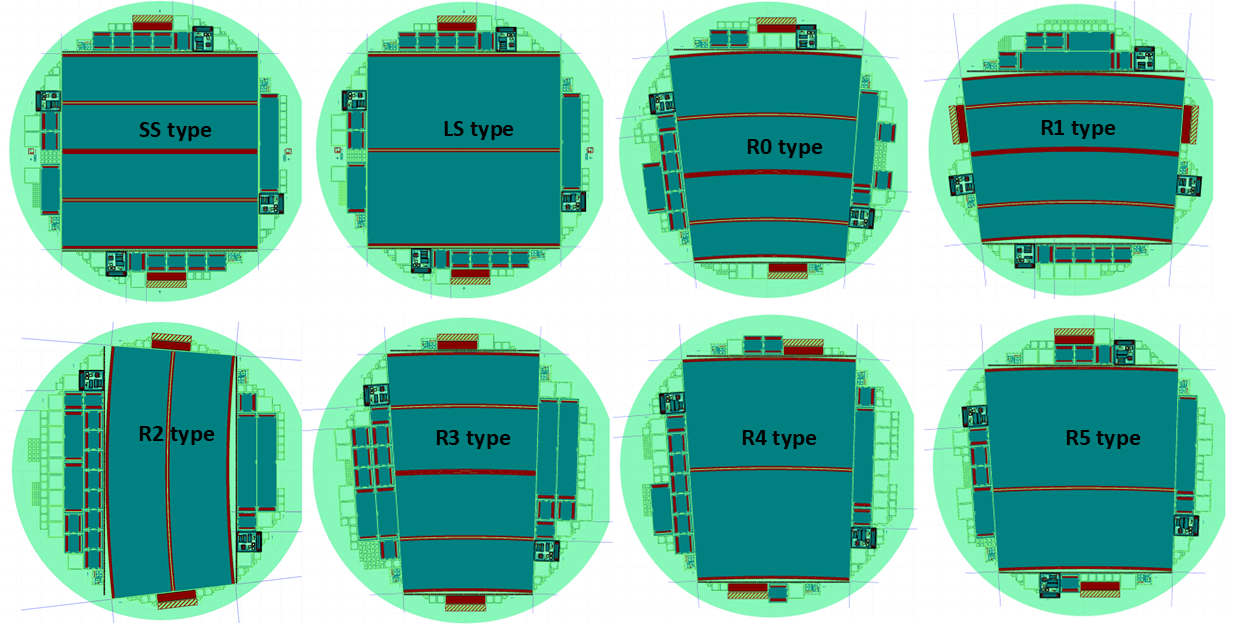}
	\caption{\rmfamily Wafer with eight ATLAS18 sensor geometries: SS and LS types for the barrel region, and R0--R5 types for the endcap regions.}
	\label{FIG:1}
\end{figure*}

The ATLAS Inner Tracker (ITk) upgrade aims to replace the existing Inner Detector with a new all-silicon system~\cite{1} capable of withstanding the extreme radiation environment expected at the High Luminosity Large Hadron Collider (HL-LHC). This upgrade involves the production of approximately 24 000 silicon strip sensors by Hamamatsu Photonics K.K. (HPK) in Japan, including both barrel for the central detector region and endcap sensors for the forward disk-shaped regions, covering 17,888 sensors to be installed and expanding the active detector area from the current 61 m$^2$~\cite{2} to 165 m$^2$ with 60 million read-out channels. The sensors are designed as single-sided, AC-coupled, n$^{+}$-in-p type, with multiple geometries~\cite{3} - two~barrel (SS, LS) and six endcap types (R0 -- R5)  corresponding to radii from inner to outer  - optimised for high-quality and precise tracking (see Fig.~\ref{FIG:1}). Each sensor is composed of two or four segments and is diced from a 6-inch silicon wafer, with the remaining ‘‘halfmoons’’ that contain various test structures used for quality assurance (QA) measurements~\cite{{4},{5},{6}}. Since the start of production in August 2021, over 98\% of sensors have been delivered by September 2025, followed by extensive quality control (QC) testing to monitor their mechanical and electrical properties. The testing is conducted at seven international laboratories (QC sites) with a monthly rate of about 500 sensors (300 sensors for barrel, 200 sensors for endcap type, respectively).

\section{QC sensor testing}
The QC testing of ATLAS18 ITk production strip sensors is a comprehensive, multi-stage process designed to monitor the quality of all fabricated silicon strip sensors~\cite{7} to ensure that their mechanical and electrical characteristics are within the specifications defined by the ATLAS collaboration. This process includes tests performed on every sensor, as well as additional tests on a sensor subset from each production batch (the batch size is between 30 -- 50 sensors), enabling monitoring of consistency across the production.

\subsection{QC tests on every sensor}
\vspace{0.2cm}
\textit{Visual Inspection:}\\
At first, all sensor surfaces, in particular their edges, are inspected manually by eye and by a microscope to identify surface defects such as scratches, chips, blotches, or other physical damage.
\vspace{0.2cm}

\textit{Visual Capture: }\\
A high-resolution image scan of each sensor's surface is obtained. This serves as a
detailed snapshot of the sensor’s condition upon arrival at the QC site. It serves as a reference against any issues that may arise during QC testing or detector
assembly.
\vspace{0.2cm}

\textit{Mechanical Bow test: }\\
The sensor’s planarity is measured using a contactless
method (e.g., a coordinate measurement machine with an optical or laser
sensor). The sensor bow is required to be less than 200 $\mu$m to ensure
sufficient flatness for module assembly (e.g., mounting electronics onto
the sensor) and subsequent loading onto local support structures.
\vspace{0.2cm}

\textit{IV and CV measurements: }\\
The bulk electrical properties of a sensor are characterised using leakage current (IV) and bulk capacitance (CV) measurements. Their key specifications are:
\begin{itemize}[noitemsep]
    \item V$_{breakdown}$ > 500 V; ensuring no breakdown during operation,
    \item I$_{leakage}$ @500 V < 100 nA/cm$^2$; indicating low defect density and good insulation,
    \item V$_{depletion}$ < 350 V; full depletion well below the maximum operating voltage of 500 V.
\end{itemize}

\subsection{QC tests on batch subset}
\vspace{0.2cm}
\textit{Mechanical Thickness test:}\\
Sensor thickness, measured either using a contactless method or with a micrometer on the corresponding halfmoon, should be (320 ± 15) $\mu$m. This measurement is required for at least one sample per batch.
\vspace{0.2cm}

\textit{Leakage Current Stability:}\\
Tested on approximately 10-20\% of sensors per delivered batch, to monitor the electric stability over time and requiring the current fluctuations to remain below 15\% during the first 24 hours. Sensors with issues from IV or Visual Inspection are prioritised.
\vspace{0.2cm}

\textit{Full Strip tests:}\\
The purpose of the full strip test is to verify the quality of the manufacturing process and the uniformity of the electrical characteristics across the wafer. Each individual sensor strip is contacted in order to identify potential defects such as metal shorts, broken implants, faulty bias resistors, or insufficient inter-strip isolation.
This test is conducted on a~batch subset, typically 2-5\% sensors per batch. The electrical specifications evaluated for each strip are:
\begin{itemize}[noitemsep]
    \item I$_{strip}$ < 200 nA, individual strip current should be less than 200 nA,
    \item 1 M$\Omega$ < R$_{bias}$ < 2 M$\Omega$, the bias resistance between metal strips and implants should be between 1 - 2~M$\Omega$,
    \item  C$_{coupling}$ > 20 pF/cm, the coupling capacitance between metal strips and implants should be higher than 20~pF/cm.
\end{itemize}
The sensor passes the test if there are:
\begin{itemize}[noitemsep]
    \item less than 1\% failed strips per segment,
    \item no more than 8 consecutive failed strips.
\end{itemize}
Sensors found during the Visual Inspection to have defects in the strip
regions are preferentially selected for this testing.
\vspace{0.02cm}

\textit{Fast Strip test :}\\
A fast strip test is a quicker version of the full test and is implemented as the default procedure instead of the full strip test at some testing sites. It is run slightly differently from a~full strip test, as it probes every fifth strip, thereby reducing the time required for each test. From these measurements, the fraction of bad strips per segment, the maximum number of consecutive bad strips, and the extrapolated number of bad strips are then estimated. Sensors that exceed the per-sensor limits in the fast strip test -- defined as a fraction of bad tested strips to be less than 1\%, estimated number of consecutive bad strips greater than 7, and fraction of extrapolated number of bad strips less than 0.5\% -- are then fully confirmed with a full strip test.

\begin{figure*}[h!]
	\centering
		\includegraphics[width=0.7\textwidth]{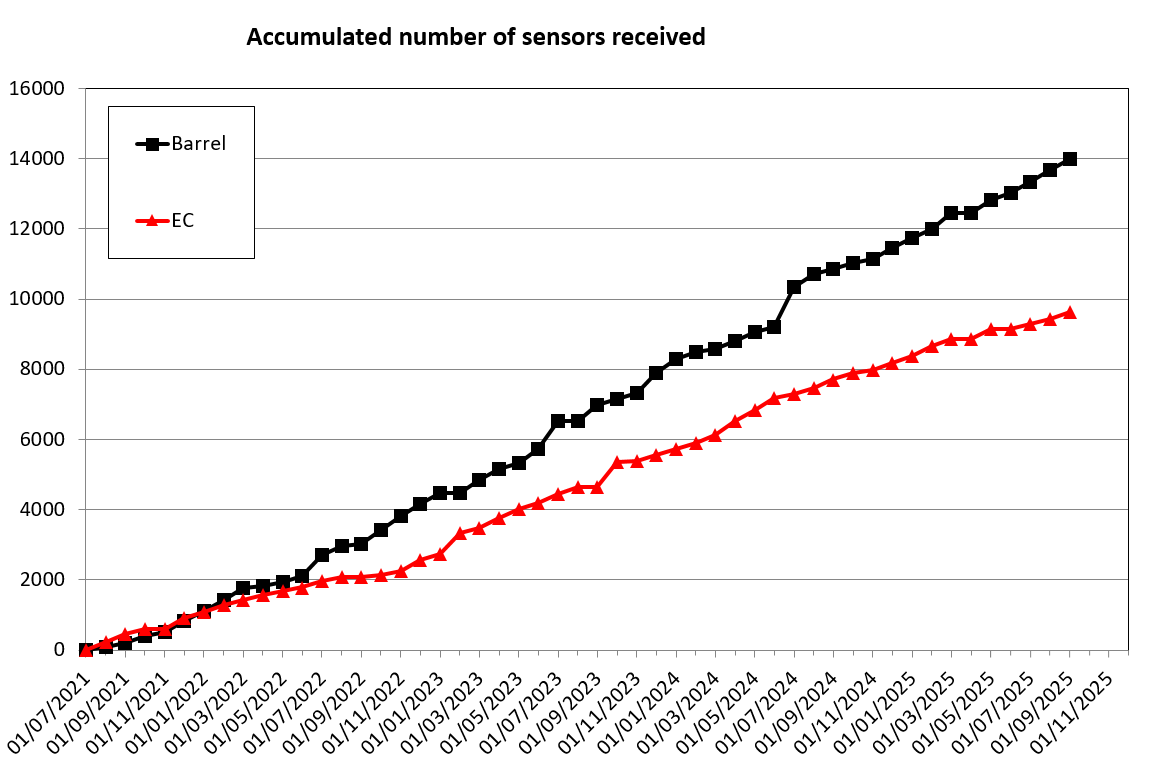}
	\caption{\rmfamily {Total number of production strip sensors received by ATLAS by September 2025. The nominal final totals are 13,987 for barrels and 9,625 for endcaps.}}
	\label{FIG:2}
\end{figure*}

\section{QC sensor properties}

The sensor QC sites have received a total of 23,614 production sensors (98\% of the required total), consisting of 13,987 barrel and 9,627 endcap sensors (Fig.~\ref{FIG:2}). Of these, 22,376 sensors have been QC tested (93\% of the required total), including 13,208 barrel and 9,168 endcap sensors, respectively. The nominal final values, excluding replacements, are 14,730 for barrel and 9,280 for endcap sensors, corresponding to a total of 24,010 production strip sensors (see Table~\ref{tab1}). Sensor deliveries and QC testing are expected to be completed by the end of spring 2026.

Distributions of the key sensor properties used in the QC evaluation are shown separately for each test type in Fig.~\ref{FIG:3}, providing an overview of the production quality and consistency across production for different sensor geometries.

\begin{enumerate}
    \item \textit{Sensor bow specs.: }< 200 $\mu$m. All sensors are well within the specification limit. The slightly broader distribution observed for the barrel sensors, compared to the endcap sensors, can be due to differences in testing setups across the testing sites (optical and laser-based systems), as well as the larger sensor size, which makes barrel sensors more sensitive to bow effects. 
    \item  \textit{Sensor thickness specs.:} (320 ± 15) $\mu$m. All measurements are also within the specification. The borderline cases were individually accepted after considering the measurement accuracy.
    \item \textit{IV leakage current normalized to 20°C at bias voltage 500 V specs.: }<~100 nA/cm$^2$. All sensors are well below the specification.

\begin{figure*}[t]
	\centering
  \begin{subfigure}{0.48\linewidth}
    \centering
    \includegraphics[width=\linewidth]{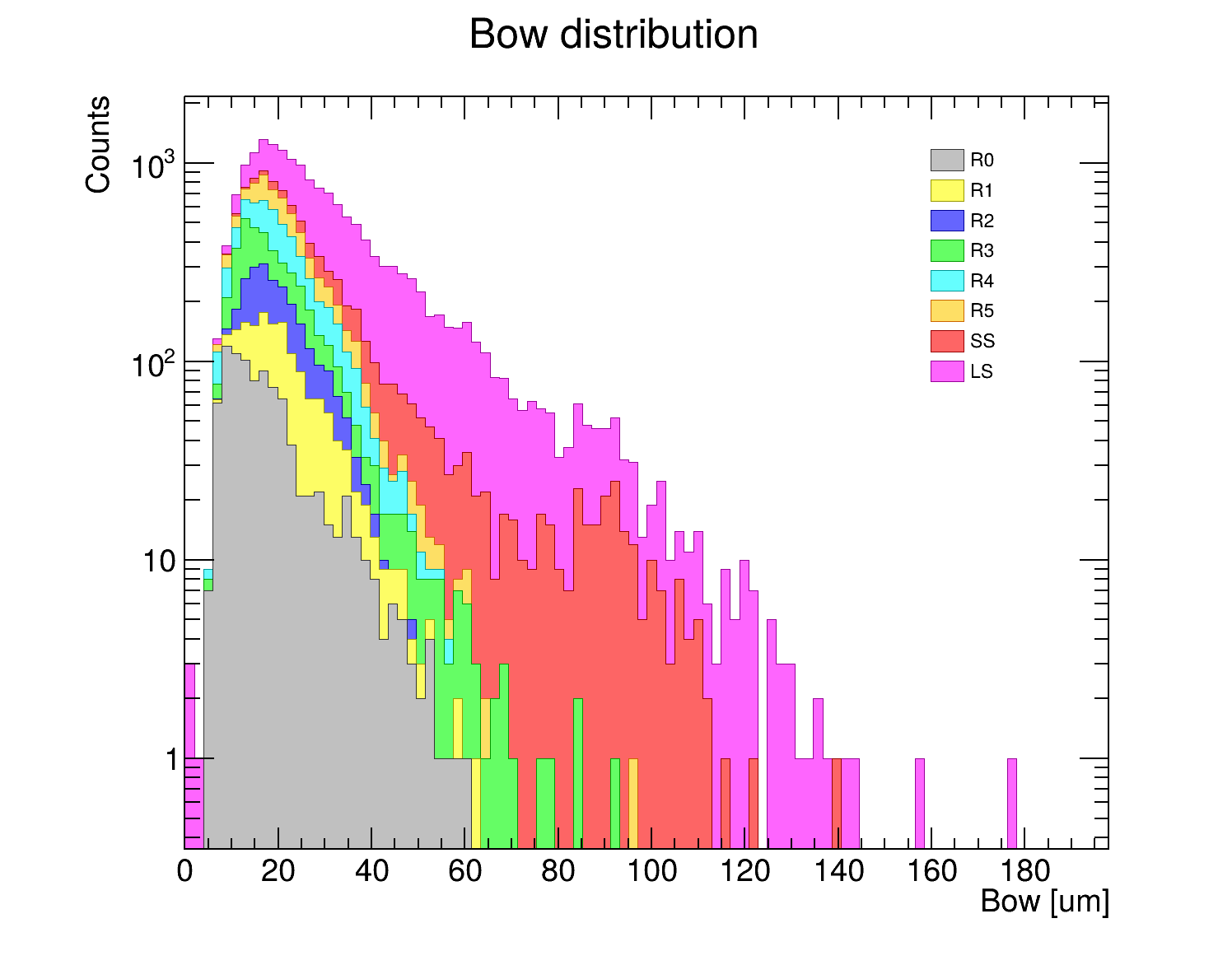}
    \caption{{Sensor bow specs.: }< 200 $\mu$m}
    \label{fig:a}
  \end{subfigure}\hfill
  \begin{subfigure}{0.48\linewidth}
    \centering
    \includegraphics[width=\linewidth]{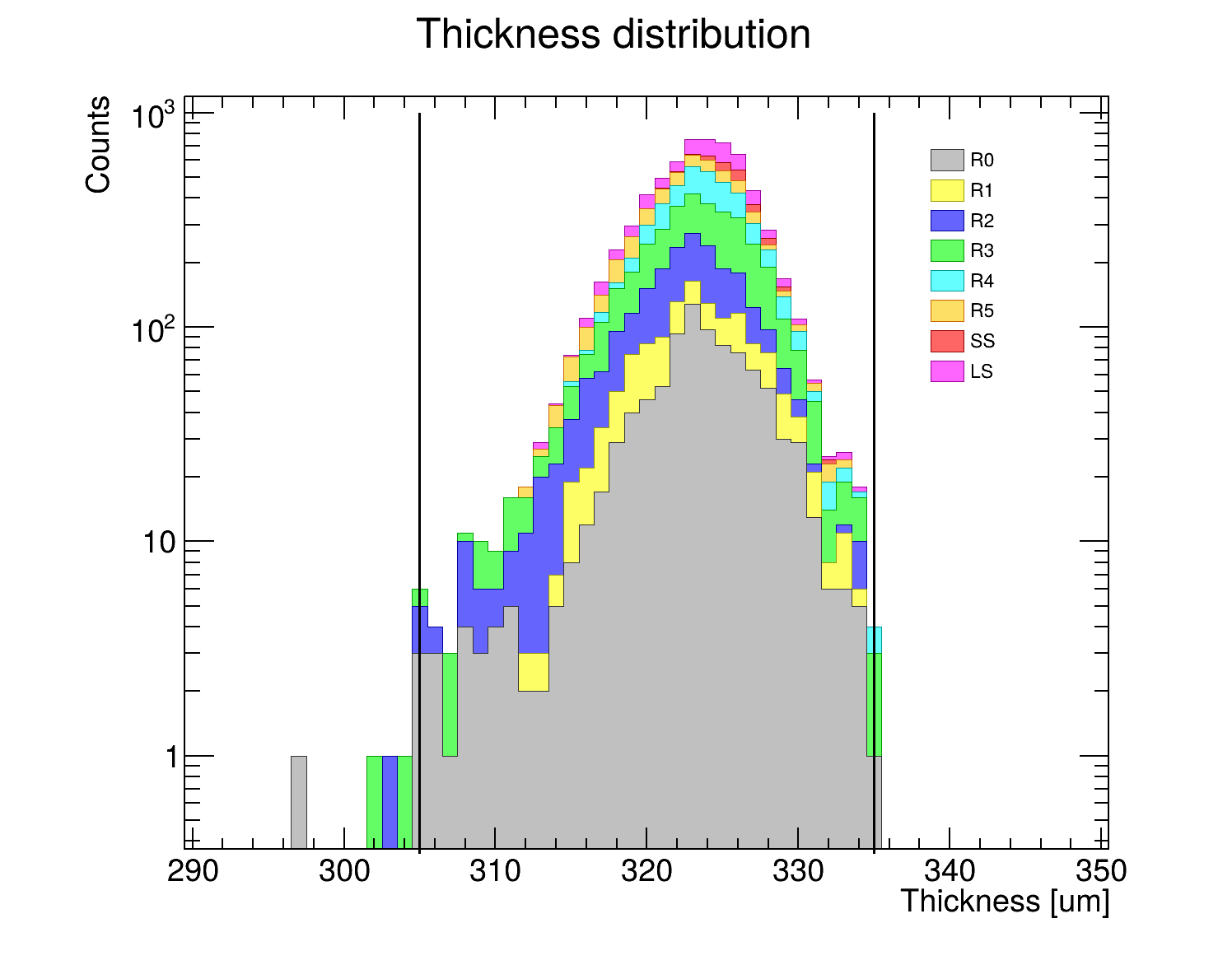}
    \caption{{Sensor thickness specs.:} (320 ± 15) $\mu$m}
    \label{fig:b}
  \end{subfigure}

  \vspace{3mm}

  \begin{subfigure}{0.48\linewidth}
    \centering
    \includegraphics[width=\linewidth]{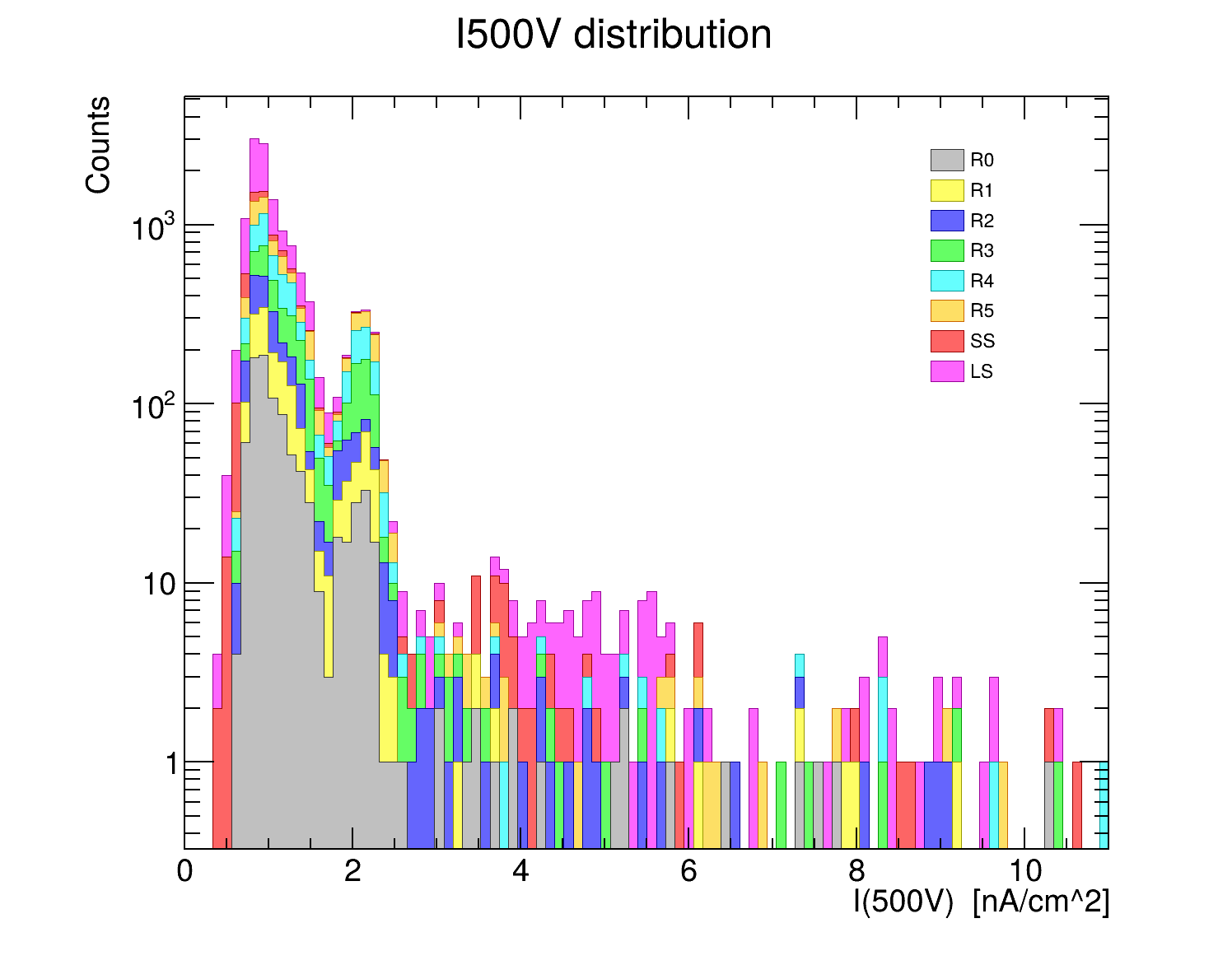}
    \caption{{IV normalised leakage current at 500 V specs.: }< 100 nA/cm$^2$}
    \label{fig:c}
  \end{subfigure}\hfill
  \begin{subfigure}{0.48\linewidth}
    \centering
    \includegraphics[width=\linewidth]{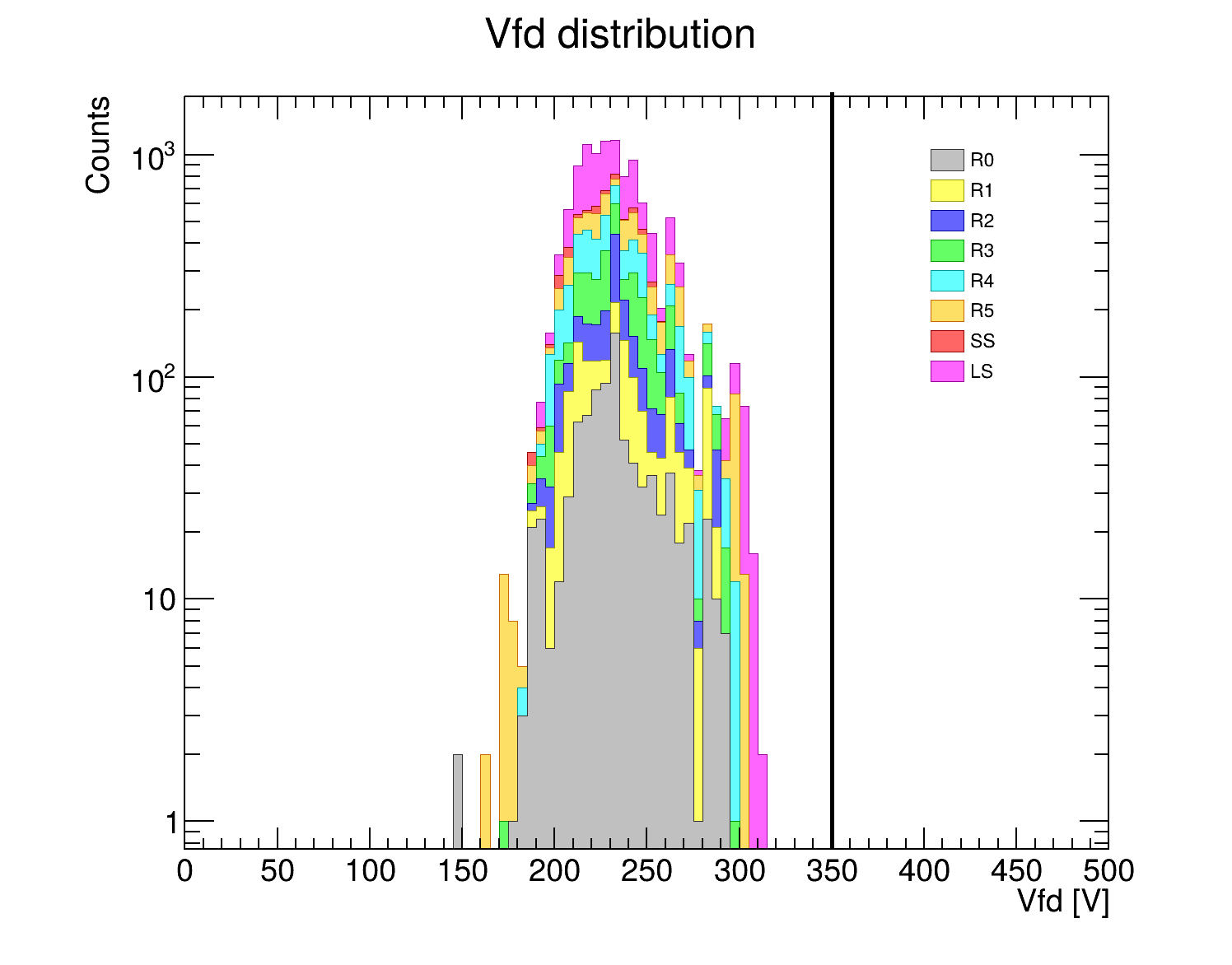}
    \caption{{Full depletion voltage (Vfd) specs.:} < 350 V}
    \label{fig:d}
  \end{subfigure}

  \vspace{3mm}

  \begin{subfigure}{0.48\linewidth}
    \centering
    \includegraphics[width=\linewidth]{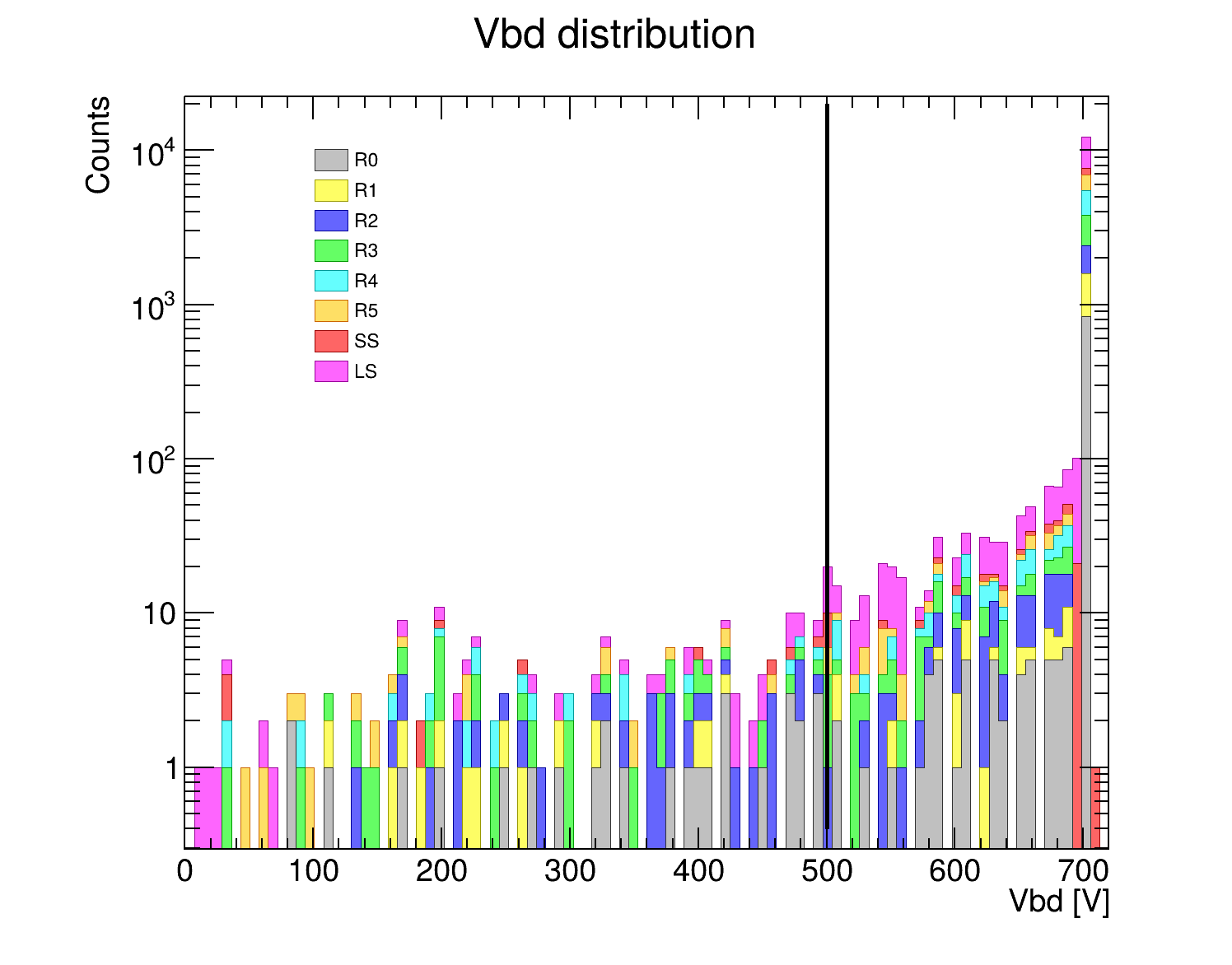}
    \caption{{The breakdown voltage (Vbd) specs.:} > 500 V}
    \label{fig:e}
  \end{subfigure}\hfill
  \begin{subfigure}{0.48\linewidth}
    \centering
    \includegraphics[width=\linewidth]{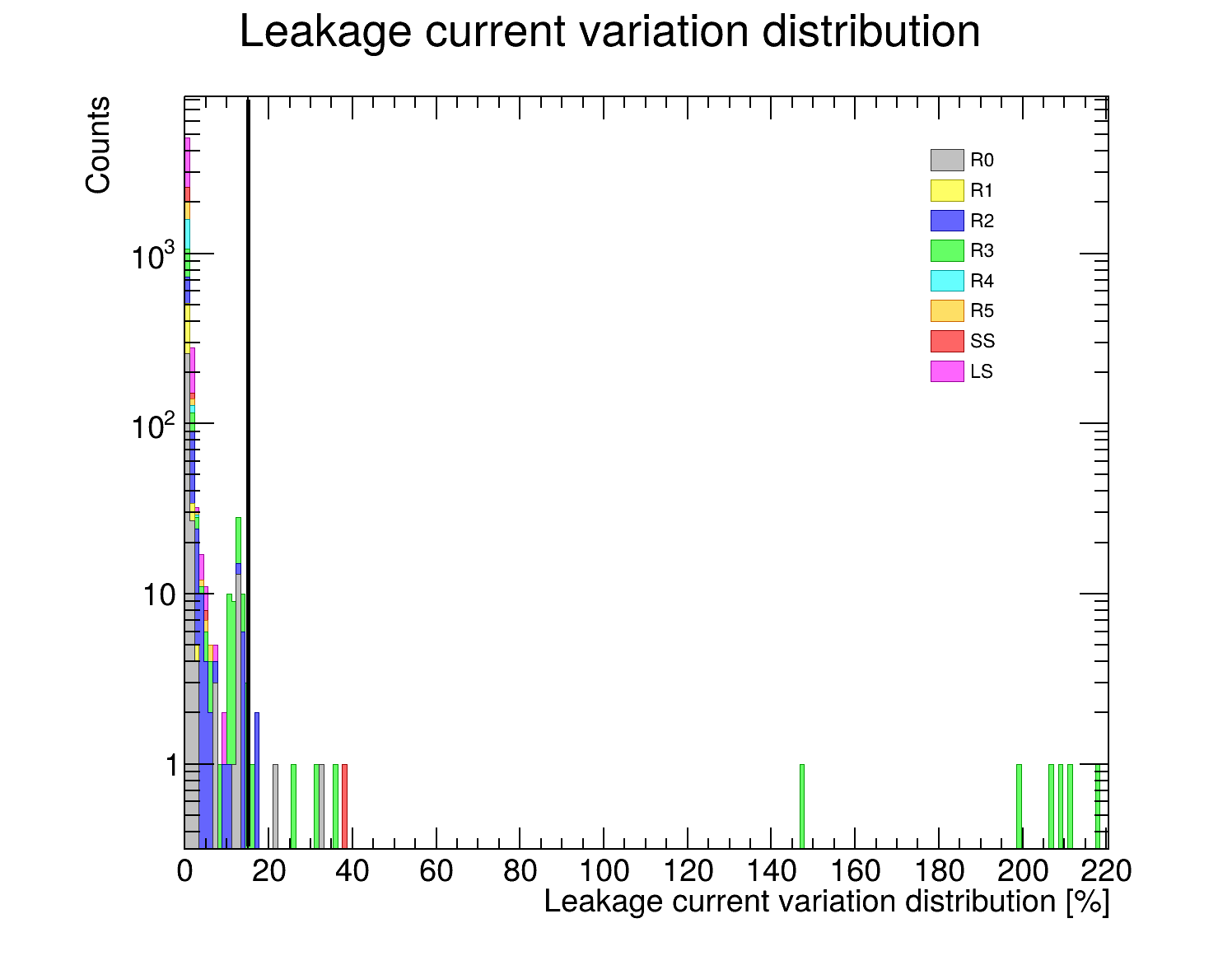}
    \caption{{Long-term leakage current variation specs.:} <15\%}
    \label{fig:f}
  \end{subfigure}

	\caption{\rmfamily Distribution of QC sensor properties per sensor type, shown on a logarithmic scale, for all measured sensors by September 2025.}
	\label{FIG:3}
\end{figure*}
\FloatBarrier
 \item \textit{Full depletion voltage (Vfd) specs.:} < 350 V. All sensors are well below the limit.
 \item \textit{The breakdown voltage (Vbd) specs.:} > 500 V. A small fraction of sensors ($\sim$1\%) did not meet the specification, while approximately 99\% passed.
    \item \textit{Long-term leakage current variation specs.:} <15\%. The majority ($\sim$99\%) of tested sensors are within this specification. Note that sensors with questionable IV behaviour have been targeted for this test.
\end{enumerate}

In addition,  Fig.~\ref{FIG:4} presents a direct comparison of the Vfd measured by HPK and by the ATLAS ITk QC sites. The two distributions show very close consistency, indicating excellent inter-laboratory agreement and confirming the robustness and reproducibility of the Vfd measurements between the vendor and the ATLAS ITk institutions.

\begin{table*}[!p]
\centering
\noindent\hspace*{2.2cm}%
\parbox{\dimexpr\textwidth-2.2cm\relax}{%
  \captionsetup{format=plain, justification=centering, singlelinecheck=false}
  \captionof{table}{\rmfamily
  Number of production strip sensors accepted and delivered by the ATLAS collaboration for different sensor geometries. \\ The accepted numbers marked with * include batch rejections.}
  \label{tab1}
}
{\rmfamily
\begin{tabular}{l|c|c|c|c|c|c|c|c|c}
\hline
\textbf{Sensor type} & \textbf{Total} & \textbf{LS} & \textbf{SS} & \textbf{R0} & \textbf{R1} & \textbf{R2} & \textbf{R3} & \textbf{R4} & \textbf{R5} \\
\hline
Nominal quantity     & 24010 & 9470 & 5260 & 1040 & 1040 & 1020 & 2120 & 2030 & 2030 \\
Delivered  & 23614 & 9470  & 4517 & 1161 & 1040 & 1109 & 2201 & 2030 & 2086 \\
\% delivered  & 98 & 100 & 86 & 112& 100 & 109 & 104 & 100 & 103 \\
ATLAS QC tested            & 22376 & 9434 & 3774 & 1040 & 1040 & 1020 & 2042 & 1996 & 2030 \\
Accepted             & 21751 & 9335 & 3712$^{*}$ & 917$^{*}$  & 1020 & 917$^{*}$  & 1939$^{*}$ & 1909$^{*}$ & 1973 \\
\% tested rejected   & 2.8   & 1    & 1.6  & 12   & 1.9  & 10   & 5    & 4.4  & 2.8  \\
\% of final accepted & 91    & 99   & 71   & 88   & 98   & 90   & 91   & 94   & 97   \\
\hline
\end{tabular}
}
\end{table*}

\FloatBarrier

\begin{table}[p!]
\centering
\noindent\hspace*{0.7cm}%
\parbox{\dimexpr\linewidth-0.7cm\relax}{%
  \captionsetup{format=plain, justification=centering, singlelinecheck=false}
  \captionof{table}{\rmfamily The most failed tests on individual sensors.}
  \label{tab3}
}
{\rmfamily
\begin{tabular}{l|c|c|c}
\hline
\textbf{Test type} & \textbf{IV} & \textbf{Vis. Inspection} & \textbf{Stability}  \\
\hline
tested & 15789& 16283 & 6069   \\
failed & 166 & 193 & 72   \\
\% failed & 1.05 & 1.19 & 1.19  \\
\hline
\end{tabular}
}
\end{table}
\section{Sensor recovery}
Upon receipt of production sensors at the QC sites, the electrostatic surface charge on both the sensors and their packaging cards is measured using electrostatic field meters. Elevated levels of this charge have been observed in a subset of production batches. A suspected source of this static charge accumulation is mechanical vibration or friction (“rubbing”) of the sensors against the packaging materials during transportation.

Furthermore, a correlation has been observed between sensors exhibiting high levels of electrostatic surface charge and an increased number of electrical test failures during QC testing.
However, a substantial fraction of such sensors that initially fail electrical tests can be recovered through the application of various treatments, which can be applied individually or in combination, including~\cite{8,9}:
\begin{enumerate}
    \item \textit{UV-A irradiation} (315–400 nm): Sensors are exposed to UV-A light for extended periods, typically 2–8 hours,
    \item \textit{UV-C irradiation} (100–280 nm): A short UV-C exposure of about 60 seconds is applied,
    \item Sensors are treated with \textit{ionising air blowers }for durations ranging from a few minutes up to~30 minutes,
    \item High-temperature exposure (“\textit{baking}”): Sensors undergo annealing at 160 °C in an oven for more than 16 hours.
\end{enumerate}
These recovery methods demonstrate that static-charge-related failures are reversible and can often be mitigated through different procedures, leading to a reduction of sensor rejection. The recovery is verified to be persistent over a period of one year rather than temporary, based on a~representative sample.

\section{Sensor rejection rate}
As mentioned, the overall sensor rejection rate is low, 2.8\%, including both individual-sensor and full-batch rejections\footnote{A batch is classified as rejected when the QA results do not meet the specifications or when $\geq$ 4 sensors fail the same QC test, indicating a possible systematic issue.}.

 In total, only six batches out of 642 delivered were rejected - two were excluded due to electrical instability with non-recoverable IV breakdown in QC, while four were rejected due to a non-uniform p-stop doping identified through  MOS capacitors and punch-through protection voltage measurements performed as part of the standard QA programme~\cite{6}. Table~\ref{tab1} summarises the acceptance and rejection numbers and percentages for the full sensor sample, as well as for individual sensor geometries.

Considering the individual sensors,  the rejection rate is only 1.8\%. The dominant causes of the individual sensor rejections are (see Table~\ref{tab3}):
\begin{enumerate}
    \item \textit{IV failures;} breakdown voltage V$_{breakdown}$ < 500 V, 1.05\% tested sensors failed this test,
    \item \textit{Visual inspection defects;} primarily surface scratches and edge chipping, 1.19\% tested sensors failed this test,
    \item \textit{Stability test failures;} leakage current variation >15\%, 1.19\% tested sensors failed this test. The higher failure rate is due to the fact that the sensors with IV issues are being targeted for this test.
\end{enumerate}

No rejections were attributed to the sensor's bow, thickness, or CV characteristics (correlated with IV failures\footnote{Every observed CV failure has been accompanied by a corresponding IV failure.}).

\begin{figure*}[!p] 
\centering
\begin{minipage}[t]{0.45\textwidth}
  \centering
  \includegraphics[width=1.1\linewidth]{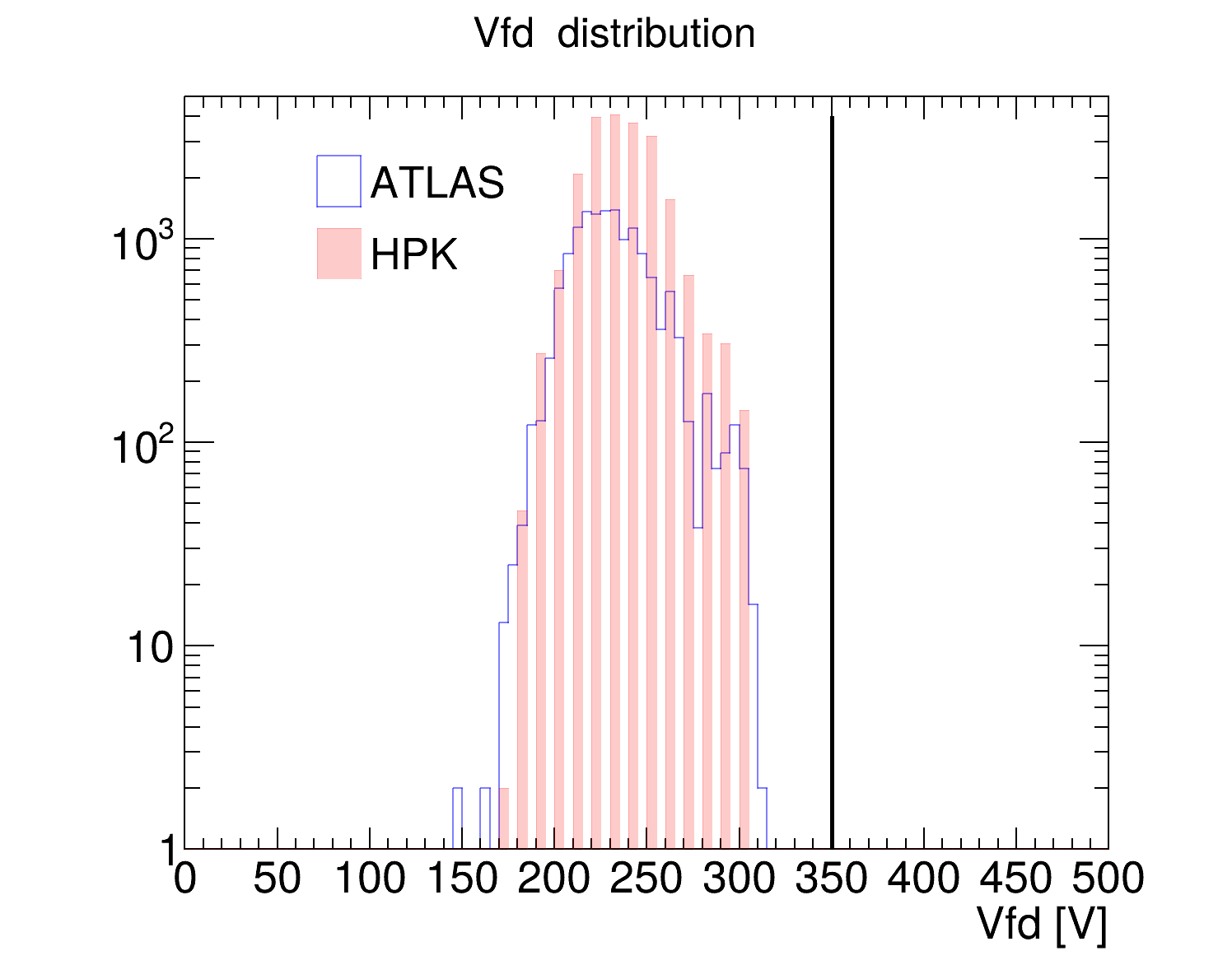}
  \refstepcounter{figure}
   \label{FIG:4}
  \parbox{\linewidth}{\rmfamily
    \hspace*{0mm}
    \textbf{Fig.~\thefigure:} Comparison of the full-depletion voltage measured\\ by HPK with that obtained by ATLAS ITk QC sites.
  }%
\end{minipage}\hfill
\begin{minipage}[t]{0.48\textwidth}
  \raggedright
  \includegraphics[width=1.05\linewidth]{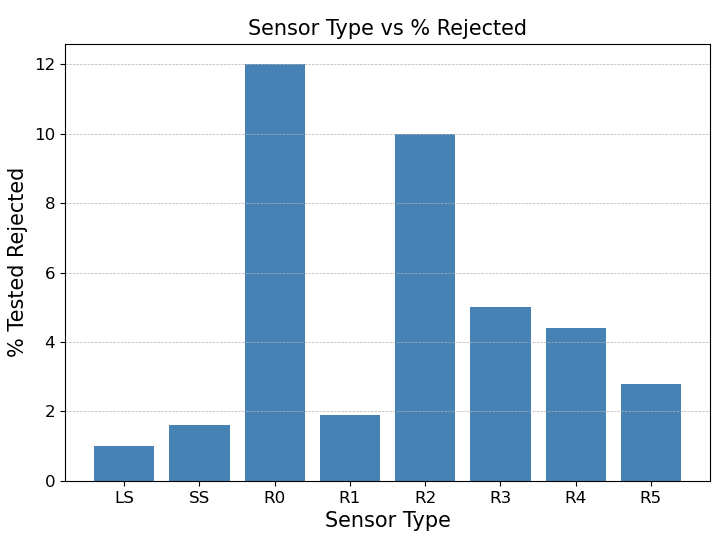}
   \refstepcounter{figure}
   \label{FIG:7}
  \parbox{\linewidth}{\rmfamily
    \hspace*{0mm}
    \textbf{\hspace{0.4cm}  Fig.~\thefigure:}{Percentage of rejected sensors per different sensor types.}
  }%
\end{minipage}
\vspace{0.3cm}
\end{figure*}

\begin{figure*}[!H]
	\centering
	\includegraphics[width=0.35\textwidth]{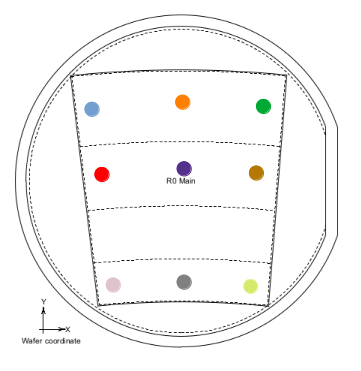}
	\caption{\rmfamily\centering{The colored markers indicate the approximate strip locations \newline \hspace{-1cm} on the main sensor where PTP measurements were performed.}}
	\label{FIG:8}
\end{figure*}

\FloatBarrier

The sensor rejection fraction shows a dependence on the sensor geometry, ranging from approximately 1\% for the LS and SS sensor types to about 10\% for the R0 and R2 sensors, including the effects of batch rejections, see Fig.~\ref{FIG:7}.

Overall, $\sim$98\% of the required total has been delivered, and 91\% of the required total has been accepted for use in the ATLAS ITk detector (Table~\ref{tab1}).

\section{P-stop doping}

Interstrip isolation is a critical performance parameter for ITk strip sensors, as it suppresses parasitic charge sharing between neighboring strips and preserves spatial resolution by limiting interstrip leakage and readout noise.

\begin{figure*}[!p]
	\centering
  \begin{subfigure}{0.95\linewidth}
    \centering
    \includegraphics[width=\linewidth]{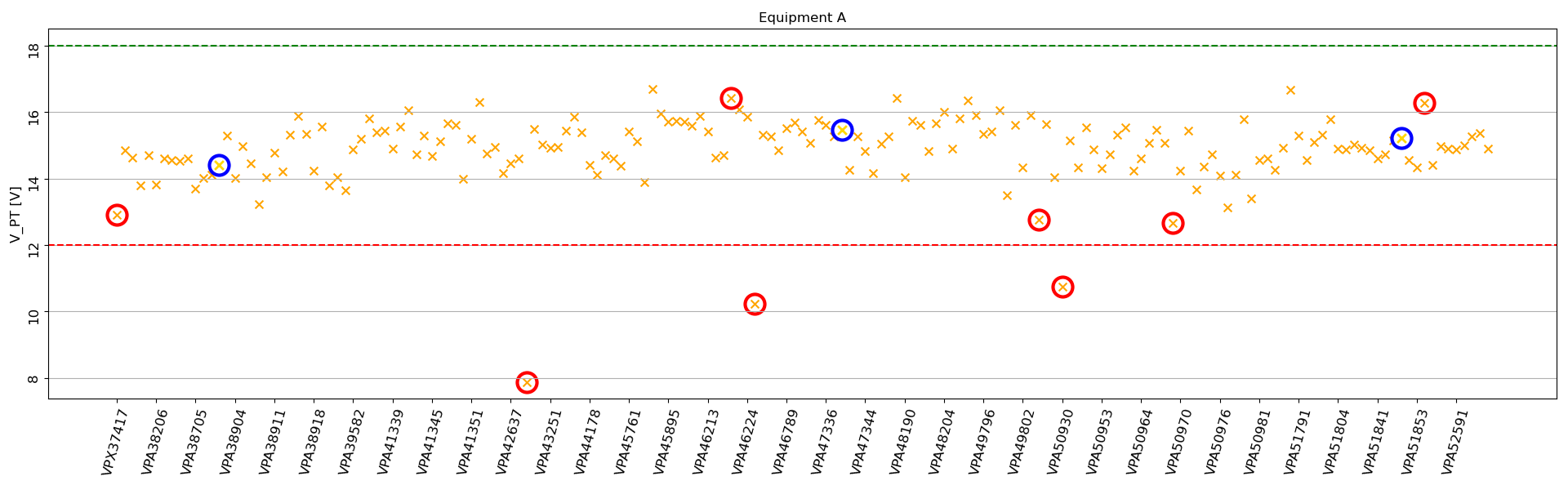}
    \caption{Selected batches for PTP measurement on main sensors from equipment A (For better visibility, only every fifth batch is mentioned in the caption.). }
    \label{fig:aa}
  \end{subfigure}
  \hfill
  \vspace{0.5cm}
  \begin{subfigure}{0.95\linewidth}
    \centering
    \hspace*{-1cm}
    \includegraphics[width=\linewidth]{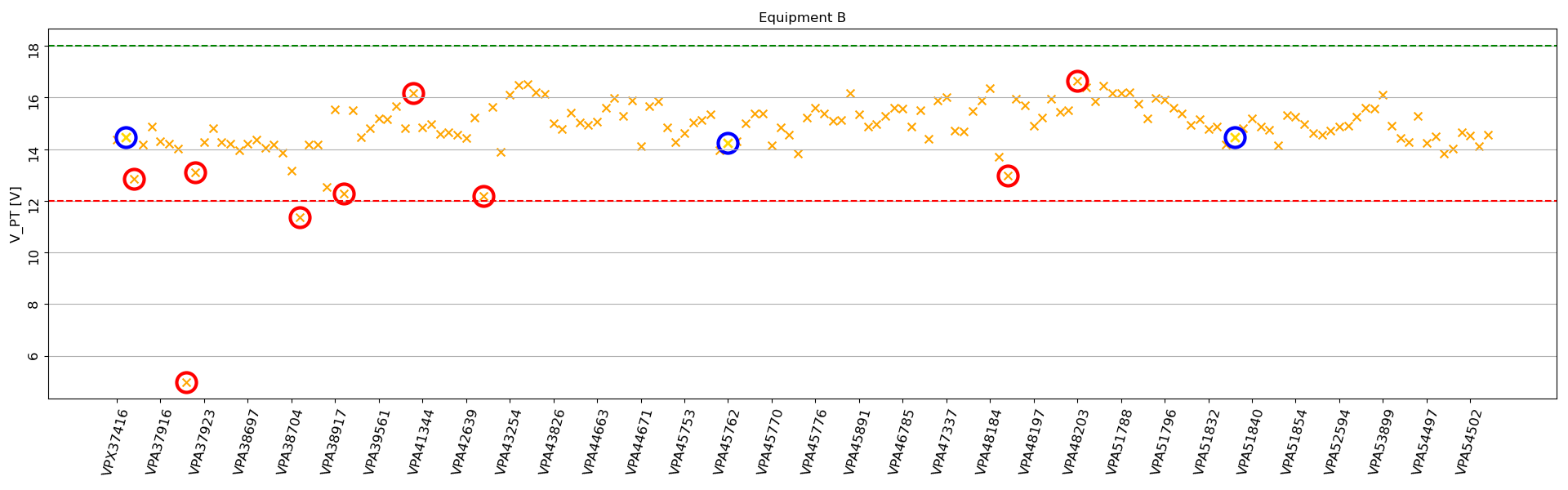}
    \caption{Selected batches for PTP measurement on main sensors from equipment B  (For better visibility, only every fifth batch is mentioned in the caption.).}
    \label{fig:bb}
  \end{subfigure}
\caption{\rmfamily Selected batches for the p-stop doping study from two different furnace processes throughout the whole production. The blue-marked batches fully satisfy the QA criteria, the red-marked batches have QA results close to the limits or out-of specifications (12 -- 18~V).}
	\label{FIG:6}
\end{figure*}

QA data indicated that the punch-through protection (PTP) threshold voltage, V$_{PT}$, is out of specification for several production batches~\cite{3,8}. The QA data are based on eight measurements per wafer, taken in a test structure on the wafer periphery, with one wafer measured per batch. To verify uniformity or non-uniformity across a wafer, dedicated PTP measurements were carried out on full-size (“main”) sensors from the affected batches.  In Fig.~\ref{FIG:8}, the colored points indicate the approximate positions at which the PTP measurements were performed to verify uniformity across a wafer. For reference, the same measurements were performed on main sensors from batches that either fully met the QA requirements (blue circles in Fig.~\ref{FIG:6}) or are close to the V$_{PT}$ acceptance limit (12 V< V$_{PT}$ <18 V; red circles in Fig.~\ref{FIG:6}). The study covers two different furnace processes, marked as equipment A and equipment B.

\begin{figure*}[!hp]
	\centering
  \begin{subfigure}{0.5\linewidth}
    \centering
    \includegraphics[width=\linewidth]{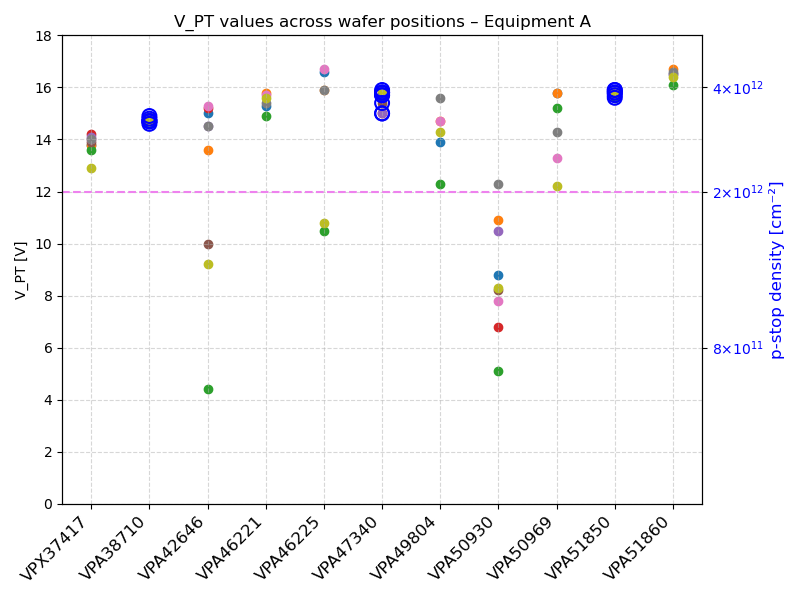}
    \caption{V$_{PT}$ values across wafer positions for equipment A. }
    \label{fig:aaa}
  \end{subfigure}\hfill
  \begin{subfigure}{0.49\linewidth}
    \centering
    \includegraphics[width=\linewidth]{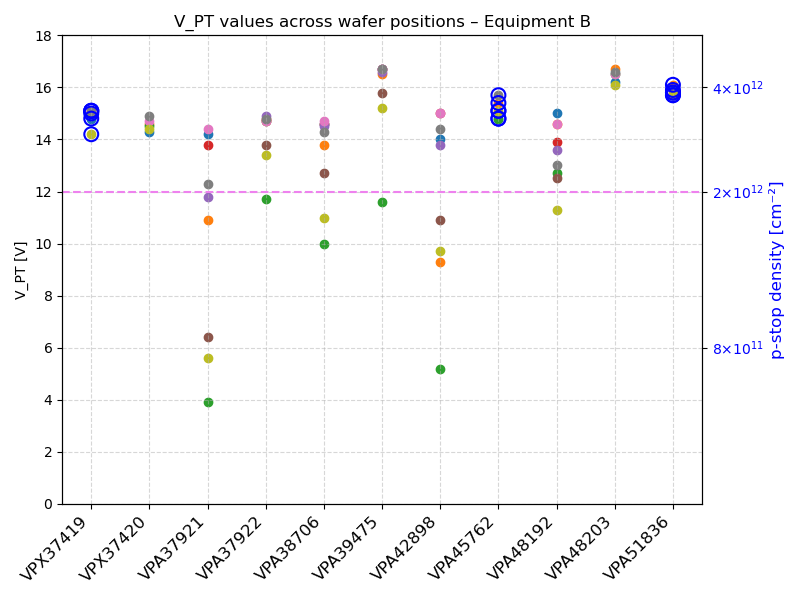}
    \caption{V$_{PT}$ values across wafer positions for equipment B.}
    \label{fig:bbb}
  \end{subfigure}
\caption{\rmfamily The punch-through voltage (V$_{PT}$) measured for selected batches for two different furnace processes and corresponding \hspace{1.5cm} p-stop density. The QA rejected batches are VPA42646, VPA46225, VPA50930, and VPA37921.}
	\label{FIG:5}
\end{figure*}

In the affected batches, V$_{PT}$ exhibits a systematic lateral gradient across the sensor: it decreases moving across the sensor surface from left to right and with increasing strip number. 
This trend is consistent with a spatial reduction of \\p-stop doping, and on the rightmost side, V$_{PT}$ of such sensors drops below the 12~V specification limit (corresponding $2\times10^{12}$ cm$^{-2}$ of p-stop density, as calculated from simulations~\cite{10}) (see Fig.~\ref{FIG:5}). Such behaviour points to deviations in the fabrication process that directly impact the electrical performance of the sensor.
\FloatBarrier

\FloatBarrier

Overall, the measurements show a clear correspondence between QA results obtained from test structures and the performance of the corresponding main sensors. Batches for which the QA parameters are comfortably within the specification limits consistently exhibit satisfactory PTP performance in the full-size sensors (Fig.~\ref{FIG:5}, where the blue circles correspond to the blue selected batches shown in Fig.~\ref{FIG:6}). QA values close to the specification limits do not necessarily result in degraded PTP performance, whereas significant QA failures are reliable indicators of poor punch-through behaviour in the corresponding main sensors. These findings underscore the critical role of QA measurements in identifying non-conforming production batches.

\section{Conclusion}
The production of silicon strip sensors for the ATLAS ITk has been progressing successfully since its start in August 2021. By September 2025, more than 98\% of the full production sensors have been delivered. The QC testing infrastructure has proven to be sufficiently robust to comfortably keep up with the delivery rate from the vendor.

The optimized QC testing workflow, combined with effective recovery procedures, has been implemented to monitor the sensor quality throughout the production, with associated minimization of sensor rejections. These efforts have significantly reduced the overall sensor rejection rate to 2.8\%, resulting in the acceptance of nearly 21,800 strip sensors for use in the ATLAS ITk detector by September 2025.
\FloatBarrier
During the production, decreasing levels of p-stop do\-ping were identified in some batches through dedicated QA measurements. Four batches were rejected due to this issue, as the punch-through protection threshold voltage dropped below the specification limit.

\section*{Acknowledgements}
This work was supported by The European Structural and Investment Funds and the Ministry of Education, Youth and Sports of the Czech Republic via projects LM2023040 CERN-CZ, and FORTE - CZ.02.01.01/00/22\_008/0004632; The Canada Foundation for Innovation under project number 36248; additional resources were provided by the Natural Sciences and Engineering Research Council of Canada; STFC grants ST/W000474/1, ST/S00095X/1, ST/X001431/1, ST/R00241X/1; JSPS KAKENHI 20K22346 and 23K13114; The US Department of Energy, grant DE-SC0010107; The Spanish R\&D grant PID2021-126327OB-C22, funded by MICIU/ AEI/10.13039/501100011033 and by ERDF/EU.

\vspace{0.2cm}

Copyright 2026 CERN for the benefit of the ATLAS Collaboration. Reproduction of this article or parts is allowed as specified in the CC-BY-4.0 license.

\bibliographystyle{cas-model2-names}

\end{document}